\documentclass[12pt]{article}
\usepackage{amsfonts,latexsym}

\newtheorem{lemma}{Lemma}
\newtheorem{cor}{Corollary}
\newtheorem{theorem}{Theorem}

\newcommand{\ct}{{T^*}'{\cal M}}
\newcommand{\hook}{\; \lrcorner \;}
\newcommand{\QED}{\mbox{} \hfill $\Box$}
\title{Stable singularities of wave-fronts in general relativity}
\author{Robert Low\thanks{email: r.low@coventry.ac.uk} \\
        Mathematics Dept,
        Coventry University,
        Coventry CV1 5FB,
        UK}
\date{}
\begin{document}

\maketitle

\begin{abstract}
A wave-front in a space-time $\cal M$ is a family of null
geodesics orthogonal to a smooth spacelike two-surface in $\cal
M$; it is of some interest to know how a wave-front can fail to
be a smoothly immersed surface in $\cal M$. In this paper we see
that the space of null geodesics $\cal N$ of $\cal M$,
considered as a contact manifold, provides a natural setting for
an efficient study of the stable singularities arising in the
time evolution of wave-fronts.\\[2ex]
PACS Numbers: 04.20C, 02.40\\[1ex]
Keywords: wave-front, contact geometry, stable singularity
\end{abstract}

Consider a smooth spacelike two-surface in a
space-time $\cal M$. Associated with this two-surfaces are two
familes of null geodesics, the ingoing and the outgoing null
congruence. Congruences of this form are called wave-fronts, and
it is of some interest to analyse the behaviour of the evolution
of these surfaces. In particular, we wish to know in what ways
the surface can fail to be smoothly immersed, at least
generically. For example, information gained from such an
analysis is relevant to the consideration of images formed
by gravitational lenses.

Analyses of the singularities of evolving wave-fronts in GR have
already been presented in the literature \cite{fs,hkp}.
This analysis differs from earlier work on the problem in that it
automatically considers variations through wave-fronts, and
requires no non-canonical choices, thereby facilitating a
geometrical understanding of the situation. It also establishes a
framework in which the application of Arnol'd's classification of
stable singularities of Legendre mappings can be applied without
the need for a great deal of intervening analysis.

We will begin by reviewing the differentiable and contact
structures of the space of null geodesics, $\cal N$, for a
strongly causal space-time, $\cal M$, and in particular for the
case when $\cal M$ is globally hyperbolic. We then go on to see
how to use the contact geometry to characterize wave-fronts and
the stable singularities arising in their evolution, making
heavy use of the analysis Arnol'd carried out for wave-fronts in
flat space \cite{arnold}.

So let $\cal M$ be a strongly causal space-time, and let $\ct$
be the cotangent bundle of $\cal M$, minus the zero section. 
Take local coordinates $(x^a,p_a)$ on $\ct$.
Then $\ct$ has two natural forms defined on it, the canonical 
one-form $\theta$ given locally by $p_a dx^a$
and the symplectic form
$\omega = d\theta = dp_a \wedge dx^a$.

In addition to these two differential forms, there are a couple
of vector fields on $\ct$ that we will need. 

The first of these is the Euler field $\Delta$, given by 
$\Delta = p_a \frac{\partial}{\partial p_a}$, which generates
dilatations on each fibre of the cotangent bundle.

The second is the geodesic flow on the (reduced) cotangent
bundle, defined by the Lorentz metric on $\cal M$ as follows:
the Hamiltonian is defined by  $H(x^a,p_a)= \frac{1}{2}g^{ab}p_ap_b =
\frac{1}{2}p^ap_a$, and the geodesic spray is the vector field
$X_G$ on $\ct$ defined by $X_G \hook \omega + dH = 0$, or, in
coordinates, $$X_G = p^a \frac{\partial}{\partial x^a} -
  \Gamma_{abc}p^a p^b \frac{\partial}{\partial p_c}$$
where $\Gamma_{abc}=g_{ad}\Gamma^{d}_{\mbox{ }bc}$ and $\Gamma^d_{\mbox{ }bc}$ is
the usual Christoffel symbol.
We note also that ${\cal L}_\Delta X_G = X_G$.

Now, to construct $\cal N$, the space of null geodesics, we
restrict our attention to $N'{\cal M}$, the bundle of (future pointing) 
null vectors, \textit{i.e.} that part of $\ct$ given by
$p^ap_b=0$ with $p^a$ future pointing. 
Furthermore, since $\Delta$ and $X_G$ are 
linearly independent and $[\Delta,X_G]=X_G$,
the two-surfaces elements defined at each point by these two
vector fields form an integral distribution.
We could consider the space of null geodesics of $\cal M$ as the
quotient manifold of leaves of this distribution.

It is sometimes more useful to consider taking the quotient by
each vector field in turn.

We could take the quotient by the Euler field, to obtain the
space of unscaled null geodesics of $\cal M$ lifted to $N'{\cal
M}$; since the Lie derivative of $X_G$ along $\Delta$ is just
$X_G$ again, although $X_G$ does not project to a vector field
on this space, it still gives a one dimensional distribution on
the resulting space. The integral curves of this distribution
form a regular distribution \cite{bc} if $\cal M$ is strongly
causal, and so the resulting space, $\cal N$ is a manifold. We
could equally carry out these two identifications in the other
order, with the same result.

It is also worth noting that under a conformal
transformation of the metric, $N'{\cal M}$ is unchanged, and
the restriction of the geodesic spray, $X_G$, to $N'{\cal M}$
is simply rescaled, so that all the geometry we are considering
is conformally invariant.

As a notational convention, we will use lower-case Greek
letters to represent points of $\cal N$ and the corresponding
upper-case Greek letters to represent the corresponding
subset of $\cal M$.

\begin{lemma} {\em
If $\cal M$ is a globally hyperbolic space-time,
and $\cal S$ is any smooth Cauchy surface for $\cal M$, then
$\cal N$, the space of null geodesics of $\cal M$ is
diffeomorphic to $T^*_1{\cal S}$, the cotangent sphere bundle of $\cal
S$.\\
\textbf{Proof} Give $\cal S$ coordinates, $x_\alpha$, where 
$\alpha$ runs from 1 to 3, and let $h$ be the Riemannian metric
on $\cal S$ induced by $g$, the Lorentz metric on $\cal M$.
Then a unit covector at $x$ in $\cal S$ can be identified with a future
pointing codirection at $x$ in $\cal M$, by adding the unit future
pointing timelike conormal to $\cal S$ at $x$, and then identifying
this covector with all its positive multiples. Thus $T^*_1{\cal S}$
may be identified with the restriction to $\cal S$ of $N'{\cal M}$.

This gives us a bijection from $\cal N$ to $T^*_1{\cal S}$, by
mapping any null geodesic $\gamma$ in $\cal N$ to $(x,v)$ where
$x$ is the point at which $\gamma$ meets $\cal S$ and $v$ is the
unit covector in $T{\cal S}$ associated with the future pointing
cotangent to $\Gamma$ at $x$. It remains only to show that this
bijection is a diffeomorphism. However, this now follows
immediately from the fact that $T^*_1{\cal S}$ gives a slice of the
foliation of $N'{\cal M}$ provided by the null geodesics, and
thus defines the differentiable structure of $\cal N$ \cite{bc}.
\QED
}
\end{lemma}

Given two Cauchy surfaces, ${\cal S}_1$ and ${\cal S}_2$, the
geodesic flow on $\ct$ defines a function 
$f_g:T^*_1{\cal S}_1 \rightarrow T^*_1{\cal S}_2$ by mapping the
point $(x_1,v_1)$ to the point $(x_2,v_2)$ where the null
geodesic through $x_1 \in {\cal S}_1$ with cotangent specified 
by $v_1$ meets ${\cal S}_2$ at $x_2$ with cotangent specified by
$v_2$. 

\begin{cor} {\em
If ${\cal S}_1$ and ${\cal S}_2$ are any Cauchy surfaces of $\cal
M$, then $f_g:T^*_1{\cal S}_1 \rightarrow T^*_1{\cal S}_2$
is a diffeomorphism.
\QED
}
\end{cor}

We can now see how the canonical form on $\ct$
gives a contact structure on $\cal N$.

To this end is convenient to list some relationships between the
Euler and geodesic vector fields and the canonical and
symplectic forms.
\begin{eqnarray*}
\Delta \hook \theta &=& 0 \\
\Delta \hook \omega &=& \theta \\
X_G \hook \theta &=& 2H \\
X_G \hook \omega &=& -dH \\
{\cal L}_\Delta \theta &=& \theta \\
{\cal L}_\Delta \omega &=& \omega \\
{\cal L}_{X_G} \theta &=& dH \\
{\cal L}_{X_G} \omega &=& 0
\end{eqnarray*}

First, we restrict everything to $N'{\cal M}$. 
This is the surface $H=0$ in $\ct$, and on
this surface we observe that $X_G \hook \theta$, 
$X_G \hook \omega$, ${\cal L}_{X_Q} \theta$ and 
${\cal L}_{X_G} \omega$ are all zero. It follows that
if we simply take the quotient space of integral curves of
the geodesic spray restricted to $N'{\cal M}$,
$\theta$ will project to the
resulting space of scaled null geodesics.

But we also want to quotient out by the action of the Euler
field. $\theta$ cannot be projected to this quotient: however,
it is homogeneous, and the distribution of planes that it
defines does project. $\theta$ thus defines a 2-form on $\cal N$
only up to scale. Denote this object by $[\theta]$.
The distribution of planes defined by the vanishing of
$[\theta]$ gives a geometric structure on $\cal N$ which is
well defined, even though $\theta$
itself need not be globally defined by this procedure.  We now
have to show that we do indeed have a contact structure. 

So let $S$ be a two-surface in $\cal N$. Then $S$ defines 
$\Sigma$, a two-parameter family of null geodesics in $\cal M$,
\textit{i.e.}  a three-surface $\tilde{S}$ in $\cal M$ which is
ruled  by null geodesics. This surface will, in general, have
self-intersections and singularities, but will not, in general,
be a wave-front.

\begin{lemma} {\em
The tangent plane to $S$ lies in the kernel of
$[\theta]$ at each point of $S$ iff the corresponding family of
null rays $\Sigma$ in $\cal M$  forms a wave-front, \textit{i.e.} it
comprises the outgoing (or ingoing) null congruence to some
space-like two surface $\Sigma$. \\
\textbf{Proof} The tangent place to $S$ lies in the kernel of
$[\theta]$ at each point iff $\theta$ vanishes on $\Sigma$,
which is well known to be equivalent to $\Sigma$ being orthogonal
to any space-like slice \cite{pr2}.
\QED
}
\end{lemma}

\begin{lemma} {\em
The planes so defined on $\cal N$ are precisely the contact
planes of $T^*_1{\cal S}$.\\
\textbf{Proof} The contact form on $T^*_1{\cal S}$ is simply the
restriction to $T^*_1{\cal S}$ of the canonical form on $T{\cal S}$.
A Legendre manifold of $T^*_1{\cal S}$ is therefore a manifold
$L$ such that for any point $(q,v) \in L$, the projection map
carries $v$ to a vector orthogonal to the projection of $L$.
But on identifying $T^*_1{\cal S}$ with $N$, we see that these are
precisely the wave-fronts, and so $[\theta]$ defines a contact
structure on $\cal N$.
\QED
}
\end{lemma}

\begin{cor} {\em
The mapping  $f_g:T^*_1{\cal S}_1 \rightarrow T^*_1{\cal S}_2$ 
is a contact mapping.
\QED
}
\end{cor}

Note that in the case where $\cal M$ is static, and ${\cal S}_1$
and ${\cal S}_2$ are both hypersurfaces orthogonal to the
timelike Killing vector, this is essentially a statement of
Huygens' principle; the above corollary is then a
generalization of Huygens' principle to a non-static
space-time.

\begin{cor} {\em
The Legendre sub-manifolds of $\cal N$ are precisely those
families of null geodesics which form wave-fronts.
\QED
}
\end{cor}

A particularly important example of this is when $\tilde{S}$
consists of all those null geodesics which pass through some
point, say $p$, of $\cal M$. In this case, one can 
show~\cite{low1,low2} that $S$ is a smooth $S^2$ in $\cal N$, 
called the sky of $p$ and denoted $P$. Thus a special case of
what we will consider is the evolution of the light cone of a
point.

We can now use this framework to provide the classification on
the stable singularities of wave-fronts~\cite{fs,hkp}, with a
considerable saving of labour.

\begin{theorem} {\em
Generically, wave-fronts will contain
only singular points of types $A_2$ and $A_3$.  At isolated 
instants, they will also contain transitions of type $A_4$ 
and $D_4$. \\
\textbf{Proof} This now follows immediately from the classifications
of Legendre mappings given in~\cite{arnold}.
\QED
}
\end{theorem}

Since the Legendre manifolds of $\cal N$ are precisely the lifts
of wave-fronts, we immediately see that this  resolution of
singularities is obtained by perturbation through Legendre
submanifolds of through wave-fronts, so that singularities may be
resolved by a small perturbation of the surface in which a
wave-front interesects any Cauchy surface. This provides some saving
of labour over the space-time approach, such as is provided
in~\cite{hkp}; the analysis presented therein may be related to
this one by the observation that the conical Lagrange manifolds
of $\ct$ are precisely those which project to Legendre manifolds
of $\cal N$.

Finally, we should note that although the analysis carried out
above has been presented for the case where $\cal M$ is globally
hyperbolic, this classification of stable singularities holds 
in rather greater generality.  If, now, $\Gamma$ is some
generator of a wave-front, we can consider the intersection of
some neighbourhood of $\Gamma$ in the wave-front with $\cal S$,
a small portion of spacelike surface through which $\Gamma$
passes. We continue to require that $\cal M$ be strongly causal;
first, so that $\cal N$ will retain its structure as a contact
manifold, and second, so that we can guarantee that by taking
$\cal S$ sufficiently small, no null geodesic in our
neighbourhood of $\Gamma$ will intersect $\cal S$ more than once.
Then by the same argument as before, wave-fronts are precisely
the Legendre submanifolds of $\cal N$. Furthermore, if we
consider the projection of a neighbourhood of $\Gamma$ to $\cal
S$, then the classification of stable singularities will be
exactly as before.

\noindent
\textbf{Acknowledgements}\\
It is a pleasure to thank the Erwin Schr\"odinger International
Institute for Mathematical Physics, Vienna, for hospitality 
and facilities, and in particular for providing such a congenial
environment, while some of this work was carried out.


\thebibliography{99}
\bibitem{fs} Friedrich, H \& Stewart, JM 1983 Characteristic
initial data and wavefront singularities in general relativity 
\textit{Proc. Roy. Soc. A} \textbf{385} 345--371

\bibitem{hkp} Hasse W, Kriele M \& Perlick V 1996
Caustics of wavefronts in general relativity
\textit{Class Quantum Grav} \textbf{13} 1161--1182

\bibitem{arnold} Arnol'd VI 1978
\textit{Mathematical methods of classical mechanics}
Springer-Verlag

\bibitem{bc} Brickell F \& Clarke RS 1970
\textit{Differentiable Manifolds}
Van Nostrand Reinhold

\bibitem{pr2} Penrose R \& Rindler W 1986
\textit{Spinors and space-time Vol 2} CUP

\bibitem{low1} Low RJ 1989 
The geometry of the space of null geodesics
\textit{J. Math. Phys} \textbf{30} 809--811

\bibitem{low2} Low RJ 1990 
Twistor linking and causal relations
\textit{Class. Quantum Grav.} \textbf{7} 177--187
\end{document}